\documentclass{moriond}

\usepackage{amsmath,amssymb}

\def\be{\begin{equation}}
\def\ee{\end{equation}}
\def\bea{\begin{eqnarray}}
\def\eea{\end{eqnarray}}

\newcommand{\Photo}{\includegraphics[height=45mm]{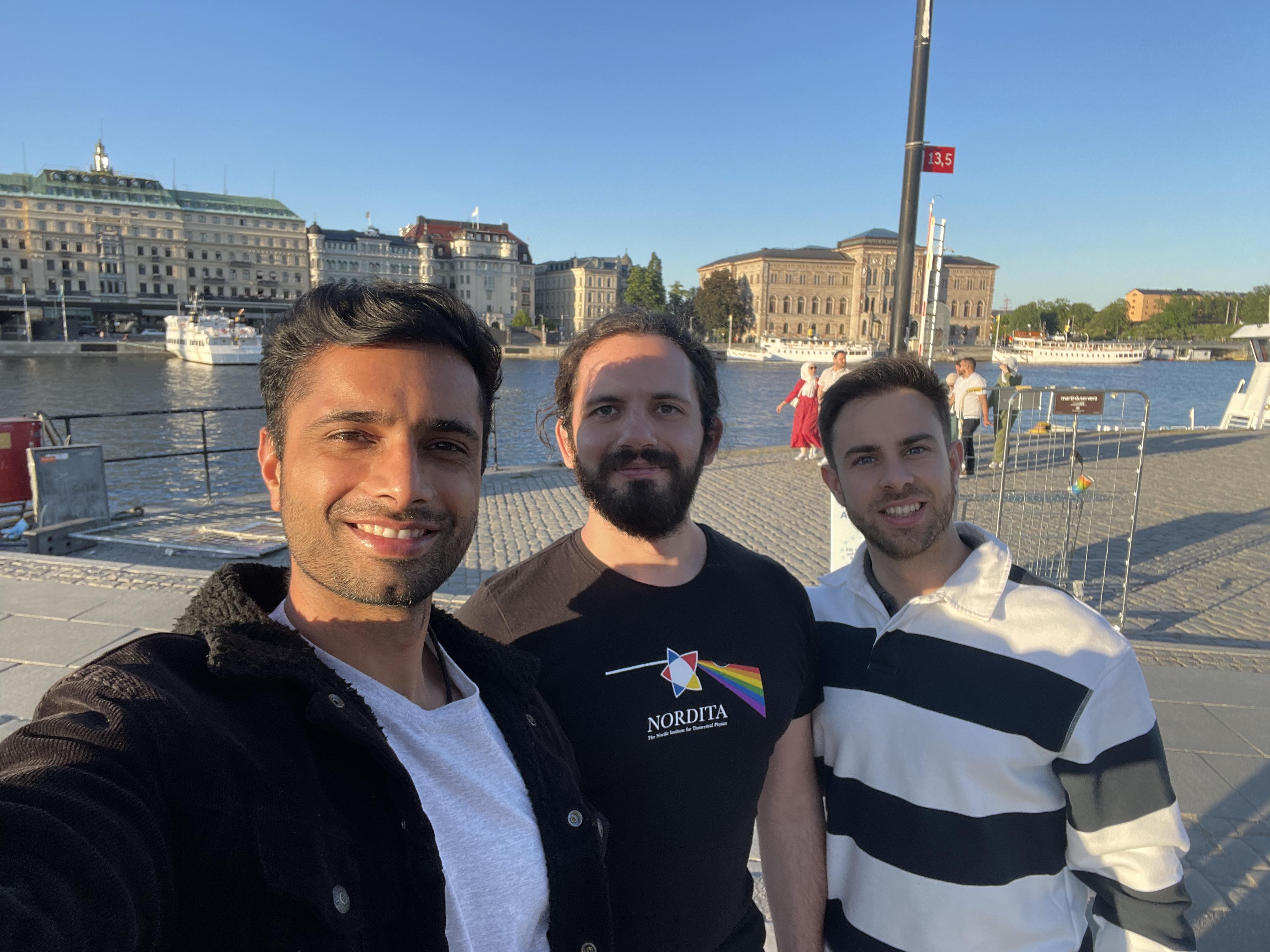}}

\begin{document}
\title{Multimessenger constraints for electrophilic feebly interacting particles from supernovae}

\author{\textbf{Pedro De la Torre Luque$^{1, 2, }$} \footnote{Speaker},
Shyam Balaji $^{3, 4}$,
Pierluca Carenza $^{5}$}

\address{$^1$ Instituto de F\'isica Te\'orica UAM-CSIC, Universidad Aut\'onoma de Madrid, C/ Nicol\'as Cabrera, 13-15, 28049 Madrid, Spain
\\ $^2$ Departamento de F\'isica Te\'orica, M-15, Universidad Aut\'onoma de Madrid, E-28049 Madrid, Spain
\\ $^3$Laboratoire de Physique Th\'{e}orique et Hautes Energies (LPTHE),
UMR 7589 CNRS \& Sorbonne Universit\'{e}, 4 Place Jussieu, F-75252, Paris, France
\\ $^4$ Institut d'Astrophysique de Paris, UMR 7095 CNRS \& Sorbonne Universit\'{e}, 98 bis boulevard Arago, F-75014 Paris, France
\\ $^5$ The Oskar Klein Centre, Department of Physics, Stockholm University, Stockholm 106 91, Sweden
 \vspace{0.4cm}}

\maketitle
\abstracts{
Several extensions of the Standard Model predict the existence of sub-GeV particles that can be copiously produced in the cores of supernovae. A broad family of these particles are dubbed feebly interacting particles (FIPs), which can have masses of up to a few hundreds of MeV. Here, we review the most recent and leading constraints on electrophilic FIPs, describing multimessenger techniques that allow us to probe the full phenomenology of the electron/positron emission produced by these FIPs; from their associated X-ray emission to the production of the $511$~keV line. Furthermore, the approach described here is independent of the specific particle model and can be translated to the coupling and other properties of a variety of different particles, such as axion-like particles, sterile neutrinos or dark photons}

\section{Introduction}
Recent ideas to understand the missing pieces or open problems of the Standard Model (SM) predict a plethora of sub-GeV feebly interacting particles (FIPs) that are connected to it~\cite{antel2023feebly} and that have been also motivated to explain anomalous cosmological and astrophysical observations (like the possible detection of a line at $3.5$~keV~\cite{PhysRevD.90.123537}, the origin of the $511$~keV line~\cite{Boehm:2003bt}, the nature of dark matter~\cite{Arbey_2021}, etc~\cite{Balaji_2023}.).
Some of the particles that have received more attention are dark photons~\cite{Fabbrichesi_2021}, sterile neutrinos~\cite{Dasgupta_2021}, the QCD axion~\cite{Peccei_2008} and axion-like particles~\cite{Choi_2021}, which have been extensively searched for either via direct~\cite{Lanfranchi_2021} or indirect methods~\cite{Raffelt:1996wa}. 

Previous studies have shown that core-collapse supernovae (SNe) are expected to produce copious amounts of FIPs~\cite{PhysRevLett.60.1797}~\cite{Caputo:2024oqc}
with masses of up to $\mathcal{O}(100)$~MeV, for typical core temperatures of $T\sim\mathcal{O}(30)$~MeV~\cite{Carenza_2024}. These have allowed us to extensively probe the production and properties of these particles either via the direct observation of SN1987A~\cite{PhysRevLett.60.1793}~\cite{Dolgov_2000} 
or from the diffuse emissions that these particles can generate~\cite{PhysRevD.105.063028}~\cite{DelaTorreLuque:2023huu}~\cite{DelaTorreLuque:2023nhh}. In fact, given the feeble interaction of these particles with the SM, FIPs can escape the SN envelope and decay in the interstellar medium, acting as continuous sources of electrons and positrons that will propagate for long durations in the Galaxy.

Here, we review the most recent astrophysical constraints on the electrons and positrons generated by FIPs produced in SNe and discuss future avenues to improve these constraints.

\section{$e^\pm$ production from FIPs in SNe and their propagation}

After leaving their parent star during SNe explosions, FIPs may decay into electron-positron pairs that populates the diffuse Galactic background. Then, these particles travel in the Galaxy following a sort of diffusive movement due to their interaction with the magnetohydrodynamic turbulence in the interstellar medium. Their propagation can be approximated by (neglecting advection)~\cite{Evoli2017jcap}
\begin{equation}
\label{eq:CRtransport}
- \nabla\cdot\left(D\vec{\nabla} f_e \right) + \frac{\partial}{\partial p_e} \left[ p_e^2 D_{pp} \frac{\partial}{\partial p_e}\left(\frac{f_e}{p_e^2}\right)\right] = Q_e + \frac{\partial}{\partial p_e} \dot{p}_e f_e \;, 
\end{equation}
where $f_e \equiv \frac{dn_e}{dp_e}$ is the density of $e^\pm$ per unit momentum. 
The term $Q_e$ is called the source term, and regulates the rate of injection of $e^\pm$ in the Galaxy. This term is usually factorised into a spatial term, that regulates the distribution of sources of these particles (in this case, featuring the spatial distribution of SNe in the Galaxy), and an energy-dependent term, that rules the spectrum of the injected particles. 

The injection of $e^\pm$ caused by FIPs has an energy distribution inherited by the parent FIP. Assuming that $e^\pm$ have an energy $E_{e}$ which is half of the decaying FIP, the injected flux can be parametrized as a modified black-body spectrum \cite{vanBibber:1988ge}
\begin{equation}
\begin{split}
    \frac{dN_{e}}{dE_{e}}&=N_{e}C_{0}\left(\frac{4E_{e}^{2}-m_{X}^{2}}{E_{0}^{2}}\right)^{\beta/2}e^{-(1+\beta)\frac{2E_{e}}{E_{0}}}\,,\\
      C_{0}&=\frac{2\sqrt{\pi}\left(\frac{1+\beta}{2m_{X}}\right)^{\frac{1+\beta}{2}}E_{0}^{\frac{\beta-1}{2}}}{K_{\frac{1+\beta}{2}}\left((1+\beta)\frac{m_{X}}{E_{0}}\right)\Gamma\left(1+\frac{\beta}{2}\right)}\,,
\end{split}
\label{eq:spectrum}
\end{equation}
where $E_{0}$ is a parameter related to the core temperature and the average energy with which FIPs are injected, $m_{X}>2m_{e}$ is the FIP mass and $\beta$ is spectral index adjusted from SN simulations. In the normalization factor, $K_{\frac{1+\beta}{2}}$ is the modified Bessel function of the second kind of order $(1+\beta)/2$ and $\Gamma$ is the Euler-Gamma function. This flux is normalized such that 
\begin{equation}
    \int_{m_{X}/2}^{\infty} dE_{e}\frac{dN_{e}}{dE_{e}}=N_{e}\,,
\end{equation}
where $N_{e}$ represents the number of electrons, equal to the number of positrons, produced in SN explosions by FIP decays, i.e.~$N_{e}=N_{e^{+}}=N_{e^{-}}$, and is the parameter from which we will set constraints. 
This general prescription enables study of the emission of $e^\pm$ from FIPs produced in SNe in a model independent way. 

Then, the rightmost term of Eq.~\eqref{eq:CRtransport} is the term representing continuous energy losses that these particles suffer while propagating. For the case of core temperatures $T\sim\mathcal{O}(40)$~MeV, injected particles will have $e^\pm$ energies around a few tens of MeV, and, therefore, the most important sources of energy losses are bremsstrahlung and ionization of neutral gas.
The leftmost term in the equation represents the diffusion term, where $D$ is the spatial diffusion coefficient, which is determined from fits of cosmic ray (CR) data at Earth (see details in Ref.~\cite{DelaTorreLuque:2023olp}). The remaining term accounts for $e^\pm$ reacceleration, with the diffusion coefficient in momentum space being directly proportional to the effective Alfven speed, $V_A$ as $D_{pp} \propto V_A^2/D$. This parameter is also determined from analyses of CR data and is key in our study, as we will discuss in the next section.

A key point to assume a stationary and smooth lepton injection is the observation that electrons and positrons propagate and interact in the galactic environment on a timescale of Gyr, extremely long compared to the SN explosions rate. Therefore, we can model the lepton injection as time-independent and smoothly following the SN distribution. The result is a diffuse flux of positrons and electrons that roughly follows the distribution of SNe.


\section{Multimessenger constraints on FIP emission}

\subsection{Local $e^{\pm}$ spectra from FIPs and constraints from Voyager-1 data}

\begin{figure}[t!]
\includegraphics[width=0.493\linewidth]{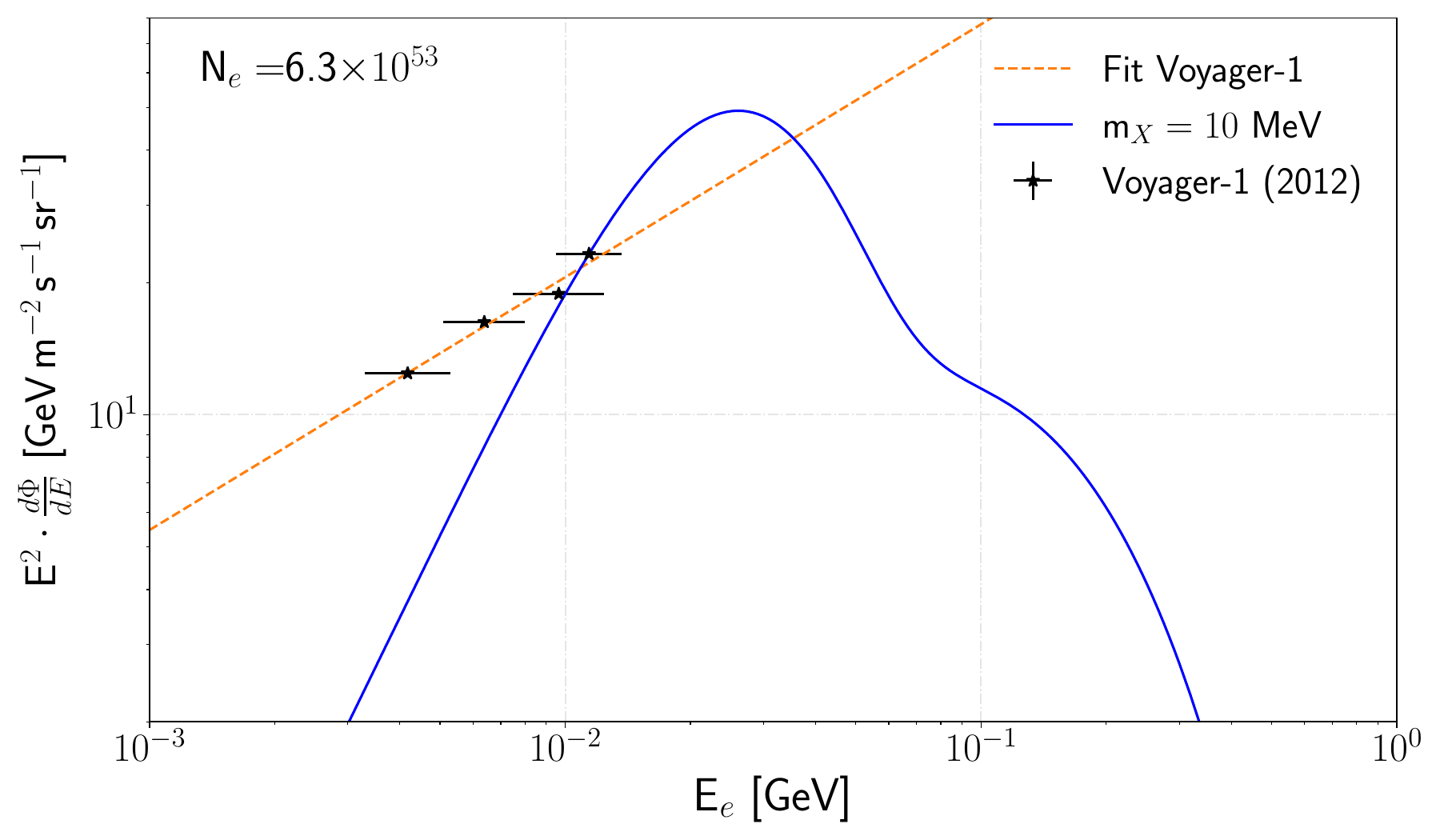} 
\includegraphics[width=0.493\linewidth]{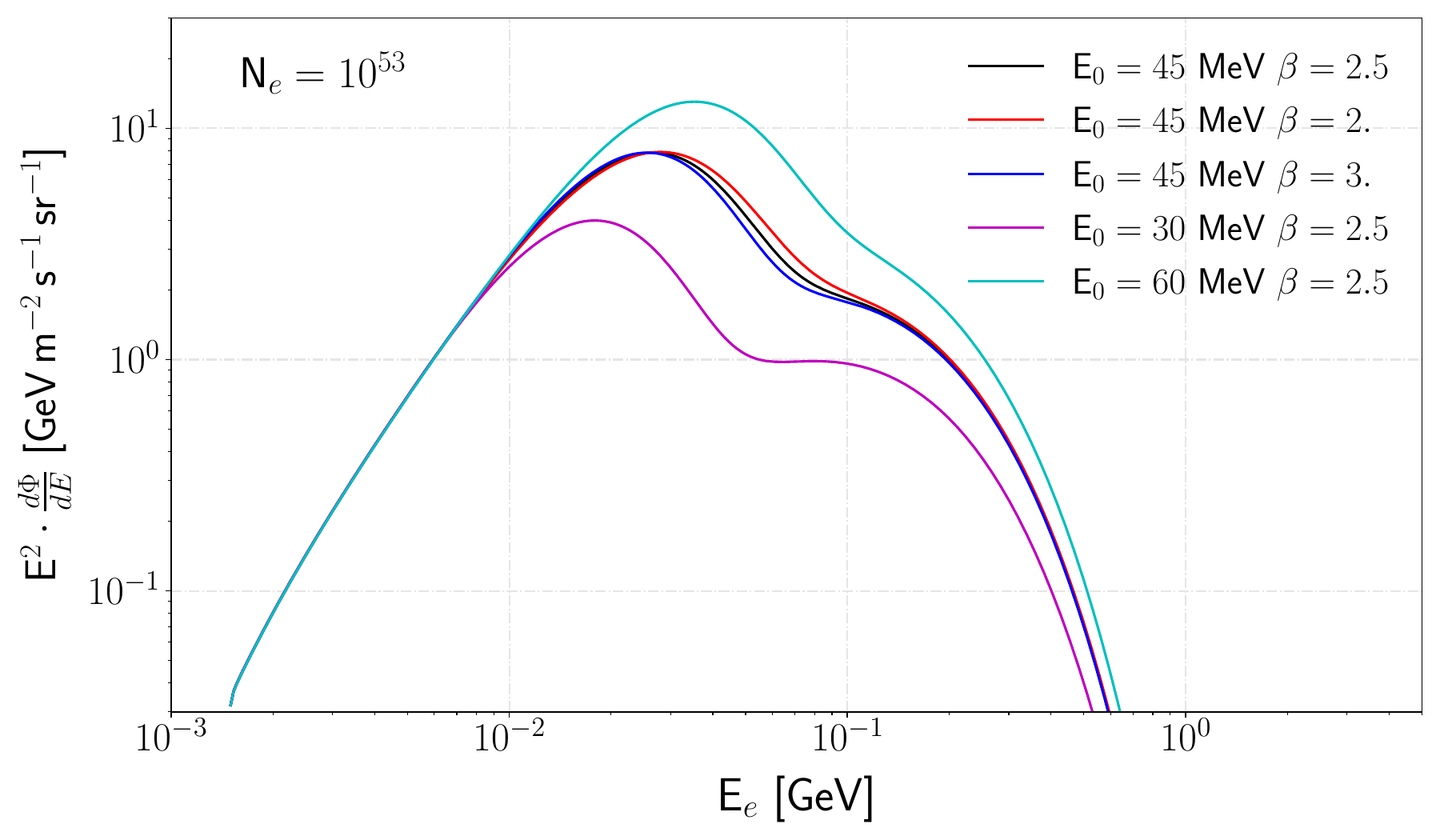} 
\caption{\textbf{Left panel}: Predicted propagated local (unmodulated) $e^{\pm}$ spectrum ($e^+$ + $e^-$) obtained from FIPs produced in SNe.  \textbf{Right panel}: Electron spectrum at the Earth location for different injection parameters, bracketing injection uncertainties. In both panels we show the case of a FIP mass $m_{X}=10$~MeV.}
\label{fig:EarthFlux}
\end{figure}

Solving Eq.~\eqref{eq:CRtransport} gives the distribution of $e^{\pm}$ and their spectrum at every region of the Galaxy. 
We show in Fig.~\ref{fig:EarthFlux} a comparison of the predicted local interstellar spectrum of electrons and positrons for a $10$~MeV FIP normalized to produce $N_e\approx6\times10^{53}$ $e^{\pm}$ per SN. we note that the propagated $e^{\pm}$ fluxes are roughly identical for FIPs with masses below $\sim30$~MeV, since this is mainly regulated by the $E_0$ parameter. This is compared to the only available CR measurements outside the heliosphere, Voyager-1~\cite{Cummings:2016pdr} data. We also show, as a yellow dotted line, a power-law fit of the Voyager-1 data, to illustrate the potential improvement that we could get on our limits.

We illustrate, in the right panel of Fig.~\ref{fig:EarthFlux}, the impact of the choice of the injection parameters $\beta$ and $E_0$ on the predicted electron/positron flux at a fixed FIP mass $m_{X}=10$~MeV and total number of injected positrons $N_{e}=10^{53}$. Typical calculations give values in the range $30$ MeV$\le E_{0}\le60$~MeV and $2\le\beta\le 3$. As we see, changing the spectral index has a very minor effect on the predicted propagated spectrum, while a change in $E_{0}$ leads to an enhancement (suppression) of the predicted flux for higher (lower) $E_0$ values, as well as a shift in the energy at which the spectrum peaks. These predictions show a second peak that is produced by the reacceleration of lower energy electrons and positrons. As shown in Fig.~8 of Ref.~\cite{DelaTorreLuque:2023huu}, the level of reacceleration drastically changes the predicted $e^{\pm}$ spectra. While in our benchmark scenario $V_A$ is set to $13.4$~km/s, values of up to $40$~km/s are reasonably found in different CR analyses, and can make this second peak to reach GeV energies and be dominant over the primary peak. However, as we see from Fig.~8 of Ref.~\cite{DelaTorreLuque:2023huu}, the level of reacceleration has negligible effect on the predicted flux at the energies covered by Voyager-1 data, making it quite a robust observable.

In conclusion, we observe that Voyager-1 provides very valuable observations to constrain electrophilic FIPs at
MeV masses limiting the number of injected number of electrons and positrons per SN to be of $N_e \simeq 6$-$9 \times 10^{53}$. The main uncertainties affecting this estimation will be the $E_0$ parameter and the distribution of SN in the Galaxy, while uncertainties on the rest of the propagation parameters affect this constraint in a minor way.

\subsection{Constraints from the 511 keV line}
\begin{figure}[t!]
\includegraphics[width=0.49\linewidth]{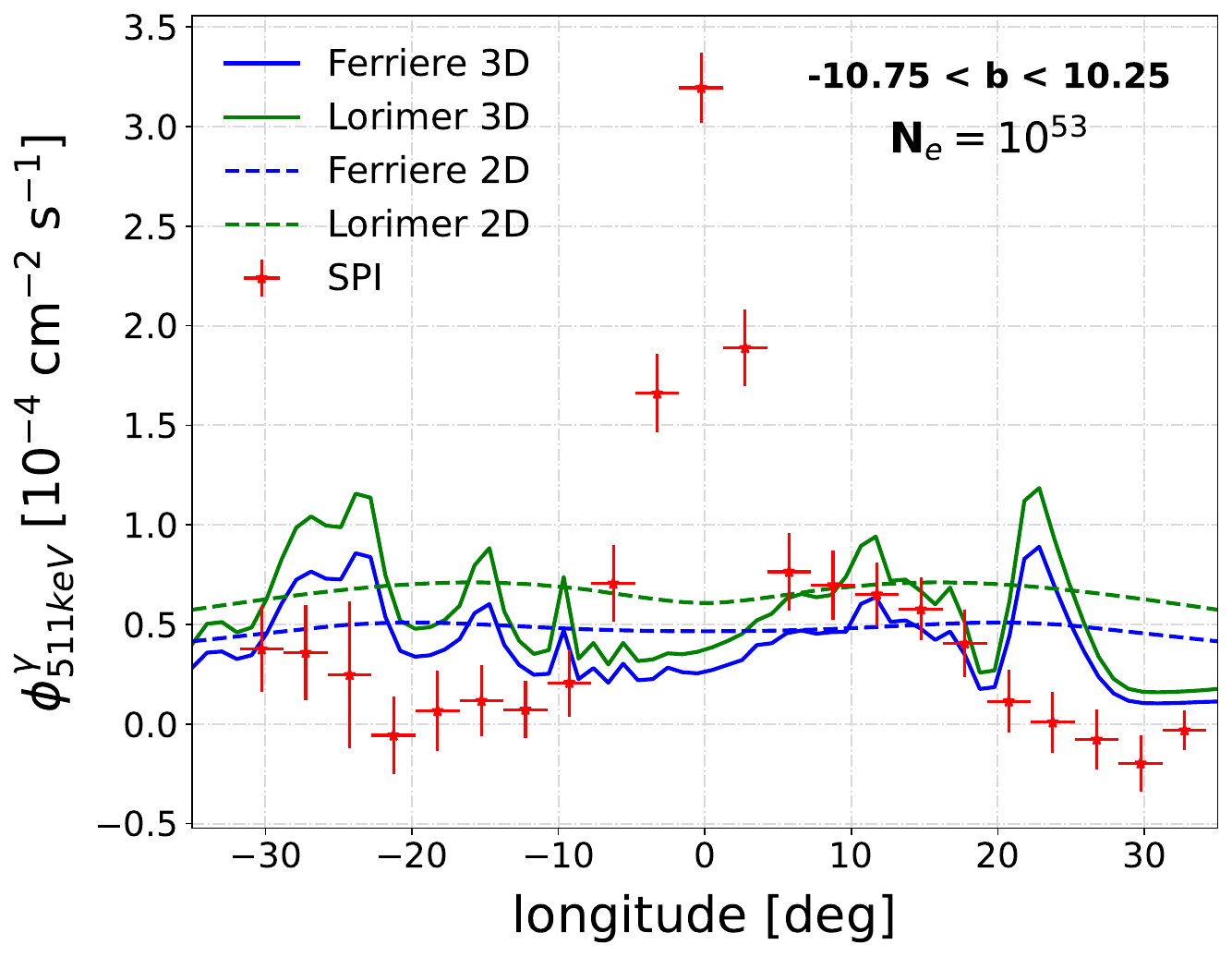} 
\includegraphics[width=0.49\linewidth]{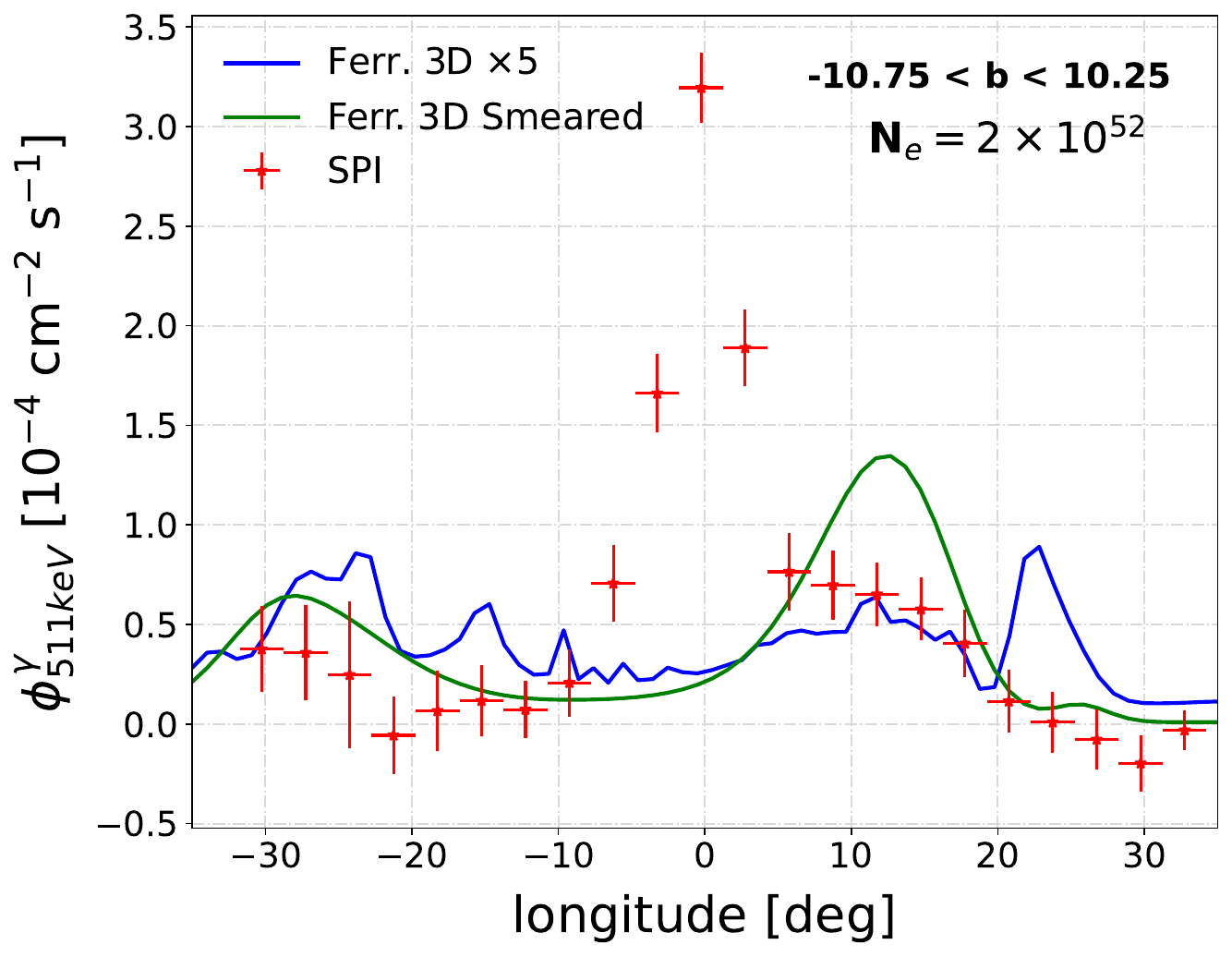} 
\caption{\textbf{Left panel}: Predicted longitude profile of the FIP-induced $511$~keV line signal with our benchmark parameters and for different SN distributions. The profiles for the 2D distributions, i.e. not considering the spiral arms structure of the Galaxy, are shown as dashed lines. SPI measurements are shown in red.  \textbf{Right panel}: Predicted $511$~keV longitude profile accounting (green line) and not accounting (blue line) for smearing of the signal emulating the effect of positron propagation. }
\label{fig:511keV}
\end{figure}

A distinctive feature indicating the presence and continuous injection of positrons in the Galaxy is the $511$~keV line. This line appears when diffuse positrons lose their energy and become thermal. After times on the order of a Myr, the thermal positrons form a positronium bound state with ambient electrons, which eventually decays (with $25\%$ chance) into two photons with energy equal to the mass of the electron i.e. $511$~keV~\cite{Guessoum:2005cb}.
Therefore, positrons injected from FIPs would contribute to the $511$~keV line emission~\cite{Bouchet:2010dj}. 
The morphology of the predicted 511~keV emission is compared to the longitudinal profile of the emission measured by SPI (INTEGRAL)~\cite{Siegert:2015knp} in Fig.~\ref{fig:511keV}. In the left panel of this figure, we see the impact of using different SN spatial distributions in our calculations, comparing also the effect of adopting the spiral arm structure of the Galaxy (``3D'' models in the legend). As we see, positron injection by FIPs leads to a roughly flat morphology for the $511$~keV line that can be a fraction of the disk emission, but cannot explain the intense emission at the central longitudes (bulge emission).

These predictions are calculated using the analytical recipe described by Ref.~\cite{Calore_2021} and~\cite{Calore_2022}, which does not model the propagation of positrons.
The right panel of that figure shows the impact in the morphology of the signal when including a smearing of the signal to account for positron propagation, which, as we see, could account for the asymmetry observed in the data (we note this asymmetry is not statistically significant).
Ref.~\cite{Calore_2022} obtained a bound from the longitude profile of the line for
$    N_{e}\lesssim 0.14\times10^{53}$.
We revisited these calculations and computed uncertainty bands on $N_{e}$ at $2\sigma$ for limits in Ref.~\cite{DelaTorreLuque:2023huu} finding an upper bound that is in the range
\begin{equation}
    0.17\times10^{53} \lesssim N_{e}\lesssim 2.50\times10^{53}\,.
\end{equation}

We remark that this does not account for systematic uncertainties in the data, which are important but difficult to estimate. In addition, the modelling of this signal and its morphology is very challenging given that different processes may occur to the thermal electrons before they thermalize. Also the effect of the electron distribution or temperature of the different gas phases is not taken into account here. 
In a recent paper~\cite{luque2024gamma}, we have performed a more detailed analysis, including the full positron propagation with state-of-the-art propagation setups. The limit on $N_e$ that we obtain is  $\simeq 3.6\times10^{52}$, with a factor of $5$ uncertainty associated to its modelling and without accounting for systematic uncertainties. 

We also remark that these results could be improved by adding a robust background model. However, there is no clear consensus on the $511$~keV emission from sources in the Galactic disk and under conservative hypotheses, there would be almost no improvement in the bounds.

\subsection{Constraints from the continuum keV-MeV diffuse Galactic photon background}

\textbf{Constraints from X-ray data: Inverse Compton emission}

Since the diffuse electron/positron population from electrophilic FIPs produced in SNe has energies above tens of MeV, their interaction with the interstellar radiation fields and gas can yield high-energy photons via inverse Compton (IC) and bremsstrahlung. While bremsstrahlung emission is more important at around a few MeV, IC emission from scattering with infrared and visible/UV radiation fields peaks at $\sim100$~keV and IC emission from the interaction with the cosmic microwave background around the keV scale (See Fig.~3 of Ref.~\cite{DelaTorreLuque:2023huu}).

\begin{figure}[t!]
\centering
\includegraphics[width=0.493\linewidth, height=0.205\textheight]{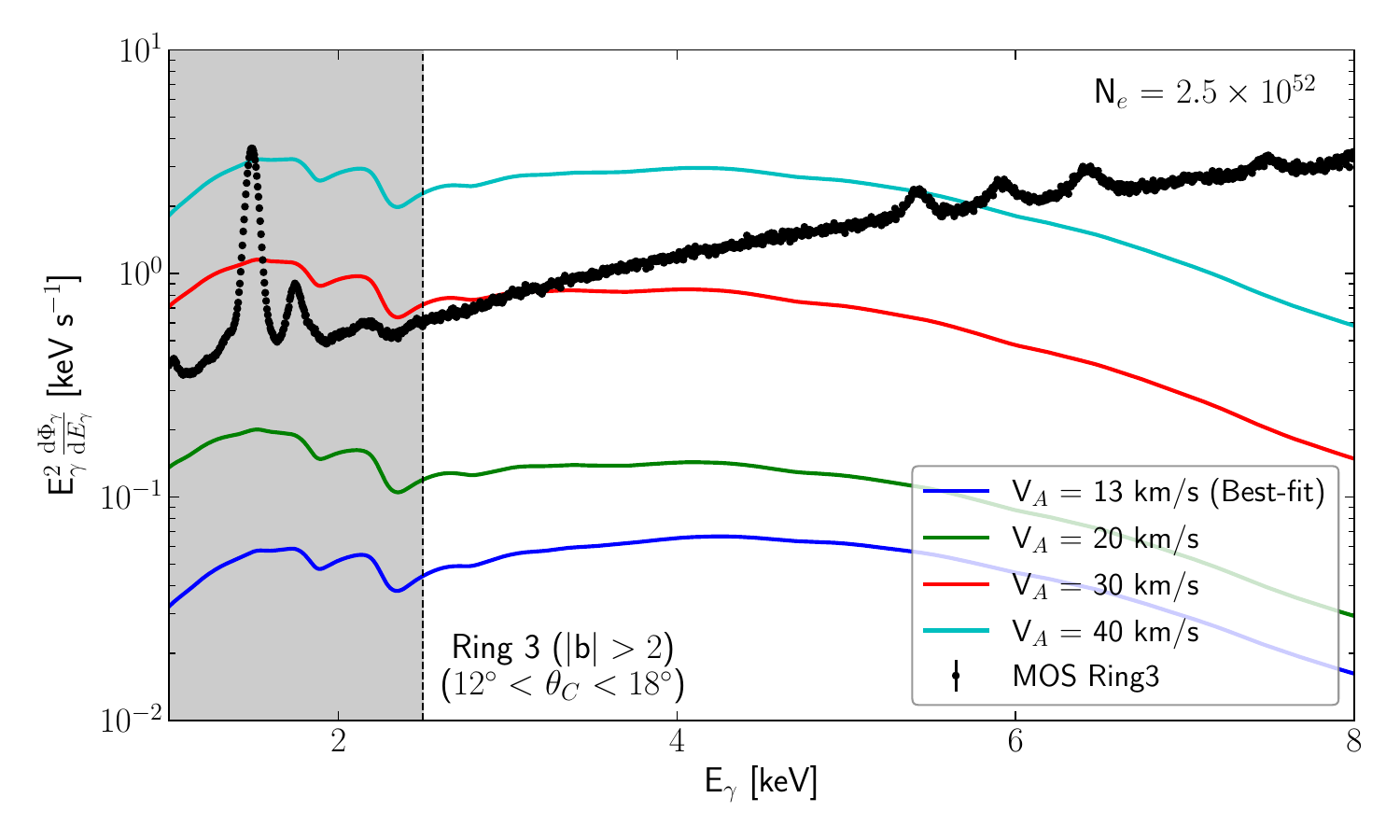} 
\includegraphics[width=0.493\linewidth, height=0.205\textheight]{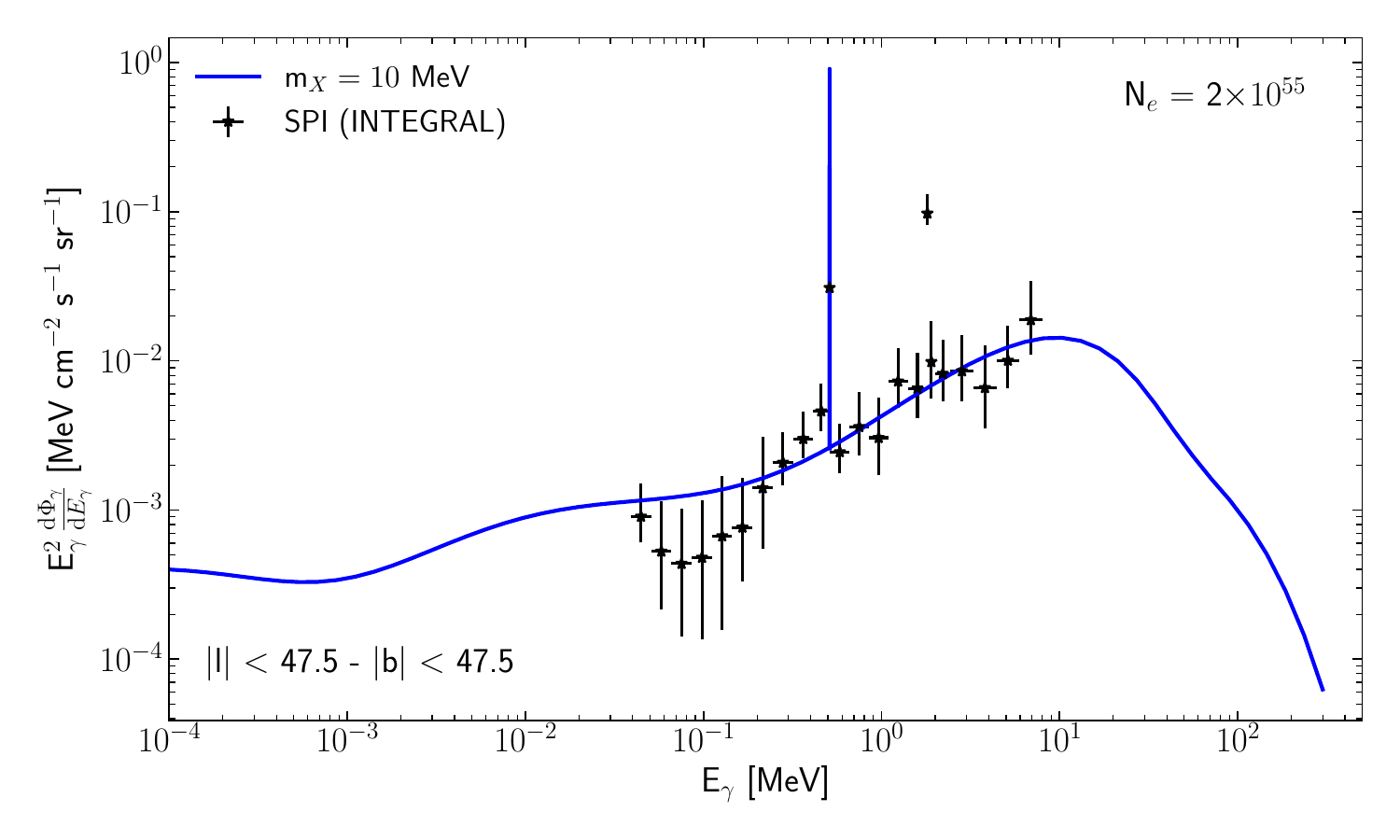} 
\vspace{0.2cm}
\includegraphics[width=0.53\linewidth]{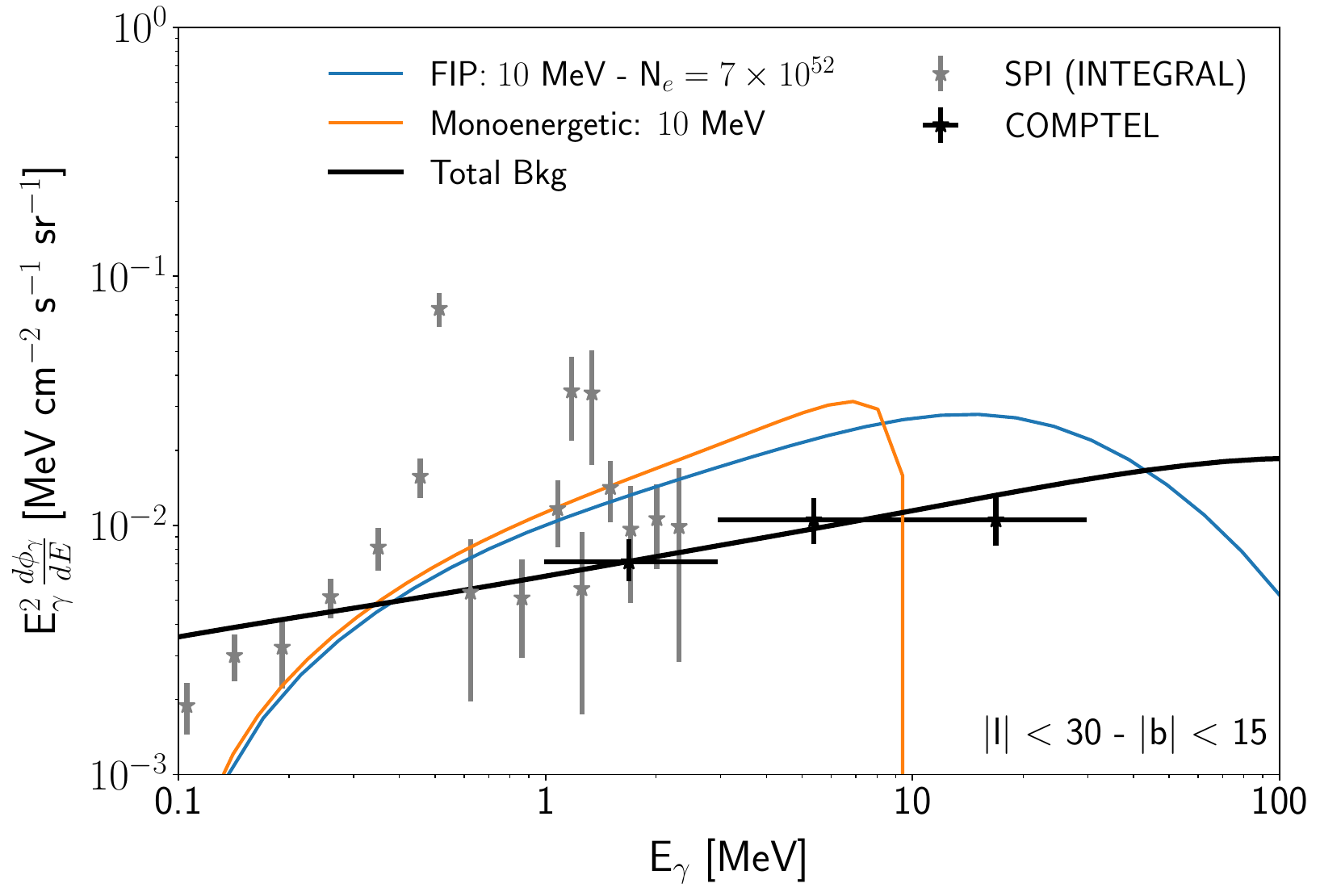}
\caption{\textbf{Top-left panel:} Comparison of MOS data for the Ring 3 with the predicted X-ray diffuse signal generated from FIP decays, for different values of $V_A$. We shade the region not used in our analyses because of high instrumental noise. \textbf{Top-right panel:} Predicted FIP-induced IC emission compared to diffuse measurements of SPI (Integral). The predicted intensity of the $511$~keV line is also shown for completeness. \textbf{Bottom panel:} Comparison of the predicted IA signals from a monoenergetic positron with $10$~MeV energy and a general $10$~MeV FIP, normalized to the integrated $511$~keV flux in the $|l|<30^{\circ}$ $|b|<15^{\circ}$ region observed by the SPI experiment.
}
\label{fig:Diffuse_Galactic}
\end{figure}

These diffuse secondary radiations are modelled using the {\tt HERMES} code~\cite{Dundovic:2021ryb}, which  performs a numerical integration of the emission along the line-of-sight at each galactic position and energy using detailed gas maps and updated ISRF models. 
Observations of the diffuse photon Galactic emission in the sub-GeV range, encompassing different zones of the Galaxy, and from different experiments have been used to set constraints on $N_e$, finding that measurements above a few tens of keV (in the so-called MeV gap), can only provide weak constraints compared to the measurements discussed above. This is due to the high instrumental backgrounds and low effective area of experiments in this energy range. Among the observations in the MeV gap, we found that COMPTEL~\cite{1998PhDT.........3K} data in a region around the center of the Galaxy, and in the range of $2$-$20$~MeV, provides the best constraints, of $
    N_{e}\lesssim 1.8\times10^{54}$.
We note that this constraint is subject to important uncertainties in the propagation setup, mainly coming from uncertainties on reacceleration, as we discuss below. In the right panel of Fig.~\ref{fig:Diffuse_Galactic}, we show a comparison of the expected diffuse gamma-ray signal generated by electrophilic FIPs with SPI measurements of the diffuse Galactic flux in the $|b|<47.5^{\circ}$, $|l|<47.5^{\circ}$ region of the sky, where we also show the predicted $511$~keV line emission in this region.

Opposite to the case of observations in the MeV gap, we found that X-ray observations allow us to strongly constrain these signals, benefiting from much lower systematic uncertainties. In particular, we use the data from the MOS detector~\cite{Foster:2021ngm} in the energy range from $2.5$ to $8$~keV in several angular rings around the center of the Galaxy.
In the left panel Fig.~\ref{fig:Diffuse_Galactic} we compare the calculated FIP signals with MOS data at the most constraining ring (ring 3). The blue line represents the signal computed with our benchmark propagation scenario, for which we can set a constraint $N_e \lesssim 2\times10^{53}$. The other colors represent the predicted signals for different levels of reacceleration (i.e. different values of $V_A$). As one can see, reacceleration has a huge impact in these signals, which translates into high uncertainties given that the value of $V_A$ is generally not well constrained, finding in different analyses values from a few km/s to $V_A = 40$~km/s (and even more, see a detailed discussion in Ref.~\cite{DelaTorreLuque:2023huu} and~\cite{DelaTorreLuque:2023olp}). 
These leaves us with a constrain from X-ray measurements that is in the range
\begin{equation}
    0.056\times10^{53} \lesssim N_{e}\lesssim 3\times10^{53}\,.
\end{equation}

As a consequence, X-ray constraints can be much  stronger than those from the $511$~keV line, suffering from less systematic uncertainties in the measurements used, but with slightly larger systematic uncertainties related to the modelling of these signals. 
We also note that current measurements at lower energies can improve the X-ray constraints by up to an order of magnitude, something that is currently being studied.

\vspace{0.4cm}
\textbf{Constraints from in-flight annihilation emission}

One of the strongest and more robust arguments against sub-GeV DM as the source of the $511$~keV emission at the bulge is that the same positrons that produce the line should produce a significant amount of diffuse gamma-ray emission due to in-flight positron annihilation emission (IA)~\cite{Beacom:2005qv}. The higher the energy of the injected positrons, the higher their chance to encounter an ambient electron in the interstellar medium and annihilate while they are still relativistic (i.e. before losing their energy and thermalizing). We have used, in Ref.~\cite{luque2024gamma}, this emission to obtain constraints on FIPs produced by SNe, since the produced positrons reach hundreds of MeV, yielding a high gamma-ray flux from IA. 

To estimate the IA emission associated with injection of positrons we follow the prescription given in Beacom and Yuksel~\cite{Beacom:2005qv}, with an adjustment to account for the fact that the signals that we study are not monoenergetic, but are emitted in a broad energy range.
The ratio of IA emission to the total emission at $511$~keV is simply determined by the energy loss rate of the positrons and the fraction of positrons annihilated via positronium states (roughly $97\%$), which is directly inferred from measurements of the line morphology.
We show a comparison of the IA emission associated with a monochromatic source injecting positrons at $10$~MeV with the expected signal produced by electrophilic FIPs created in SNe in the bottom panel of Fig.~\ref{fig:Diffuse_Galactic}, where one can see how the diffuse data from COMPTEL can provide very strong constraints on $N_e$. We also remark that the IA spectrum is quite different from the background spectrum (consisting mainly of IC emission from CR electrons), so that one could do a more detailed search for bumps in the diffuse gamma-ray emissions to better search for these signals.
We obtain a constraint on $N_e$ from IA emission of $N_e < 4\times10^{52}$ without including background emission, while the bound strengthens to $N_e < 8.9\times10^{51}$ when including it, for our benchmark propagation parameters. These bounds are uncertain by a factor of a few due to uncertainties in propagation parameters (mainly $V_A$), making IA emission an even more robust and promising way to constrain injection of positrons from any kind of exotic source.

\begin{figure}[t!]
\centering
\includegraphics[width=0.5\linewidth, height=0.193\textheight]{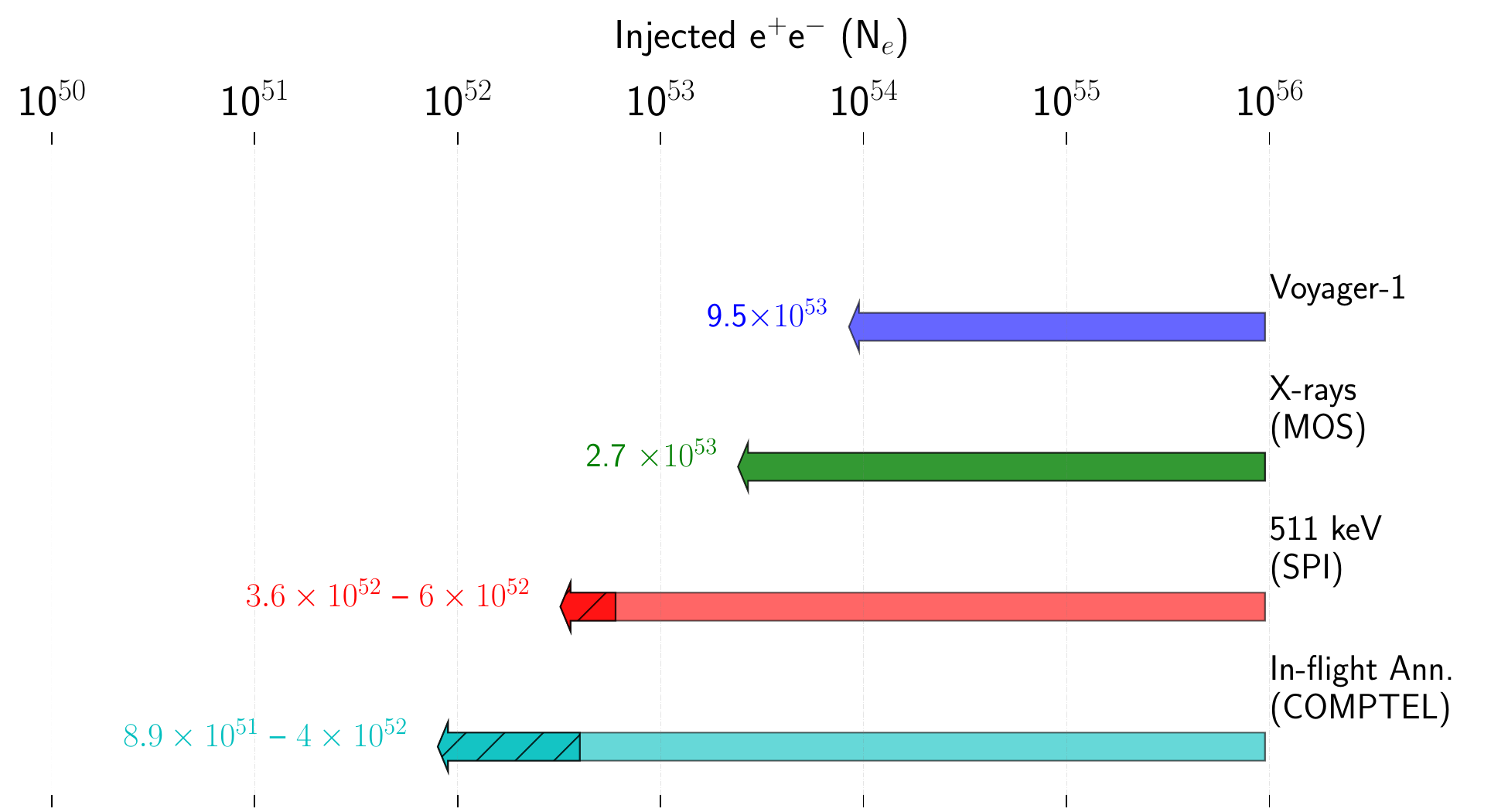} 
\includegraphics[width=0.49\linewidth]{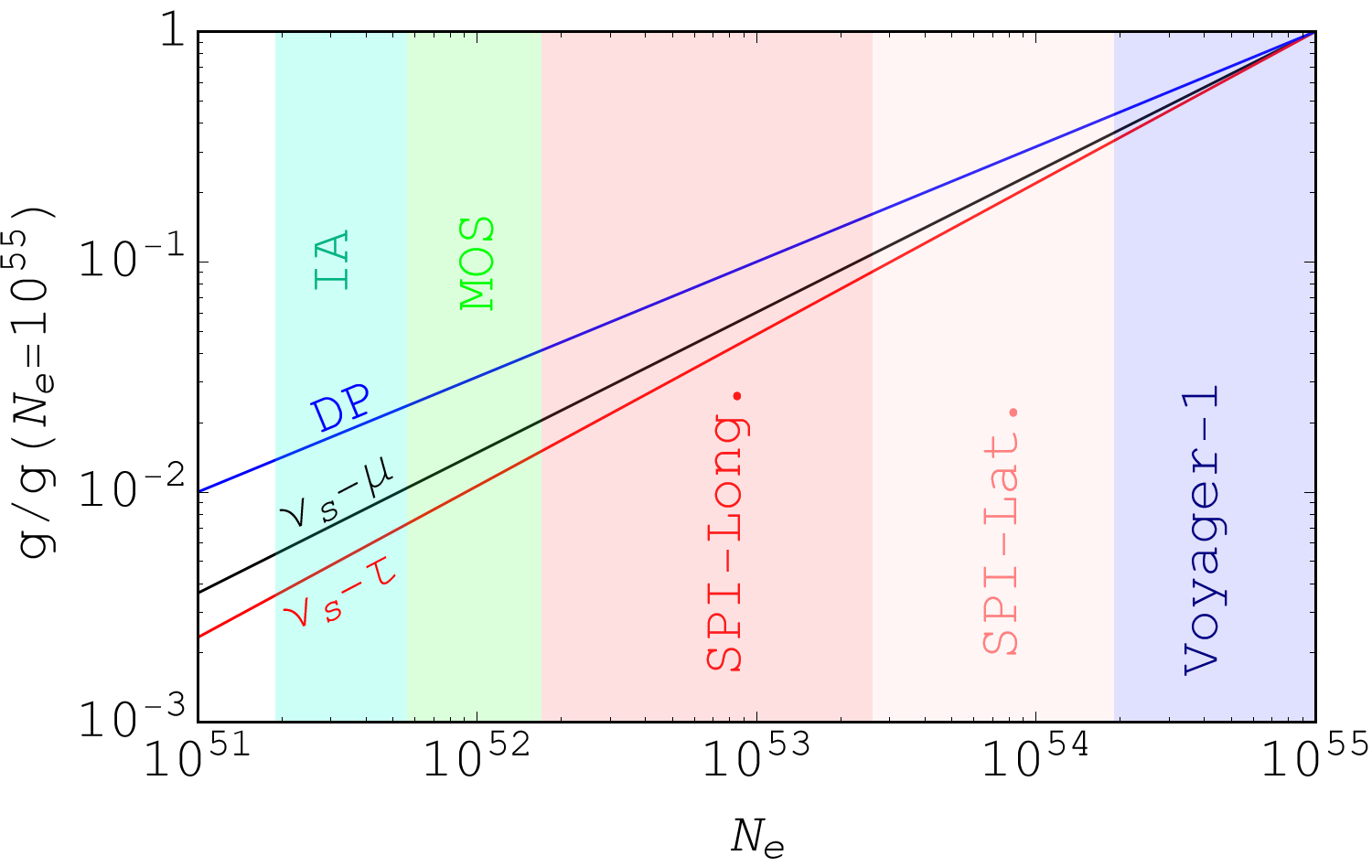} 
\caption{\textbf{Left panel:} Bounds at the $2\sigma$ level on the number of electrons and positrons injected per SN ($N_e$) for the general FIP case. We show here the limits derived from the local electron-positron flux by Voyager-1 (blue), X-ray emission from IC scattering of FIP-induced electron fluxes (green), the longitude profile of the $511$~keV line (red) and the ones derived comparing IA emission with the diffuse $\gamma$ ray observations of COMPTEL in the $|l|<30^{\circ}$ - $|b|<15^{\circ}$ region (cyan). \textbf{Right panel:} Translated bounds on $g$, for popular FIP models discussed in the text. The colored regions refer to the various bounds on the maximal $N_{e}$ (i.e. in the most optimistic case). In particular, the strongest bound is shown by the purple region region with the constraint on the diffuse Galactic gamma-ray flux due to IA emission.}
\label{fig:Lims}
\end{figure}

\section{Discussion}

We have reviewed recent and powerful astrophysical constraints on the coupling of electrophilic FIPs produced in SNe with electrons and positrons, showing that their rich phenomenology encompasses several processes in a very broad energy range. These constraints involve direct observations of electrons and positrons by Voyager-1, the $511$~keV line and the diffuse X-to-gamma-ray Galactic flux (either via IC or IA emission produced by the electrons and positrons), whose constraints are summarized in the left panel of Fig.~\ref{fig:Lims}, for our benchmark propagation setup. We note that these constraints are much stronger than those obtained from SN cooling, not discussed here, as shown by Ref.~\cite{Carenza_2024} (see also Fig.~11 of Ref.~\cite{DelaTorreLuque:2023huu}). 

An advantage of our approach is that these constraints can be obtained without assuming any specific FIP model, with the further advantage that the bound on $N_e$ can be translated into an approximate bound on the coupling constant of specific FIP models (e.g. sterile neutrinos, dark photons, axion-like particles, etc.) in the weak-mixing regime. As done in Ref.~\cite{DelaTorreLuque:2023nhh} and~\cite{DelaTorreLuque:2023huu}, the relation of $N_e$ with the coupling, $g$, of electrons to a given particle physics model can be parameterized as $g\propto N_e^{-\alpha}$. The spectral index $\alpha$ will depend on the specific particle model and can be analytically computed~\cite{DelaTorreLuque:2023huu}. In this way, we show in the left panel of Fig.~\ref{fig:Lims} how the bound on $N_e$ translates into a bound on $g$ for the case of dark photons ($\epsilon$) and sterile neutrinos coupling to muons ($|U_{\mu 4}|^2$) and taus ($|U_{\tau 4}|^2$). 

In summary, our study emphasizes the importance of Galactic signals produced by SNe in the exploration of FIPs and shows how multimessenger observations from diverse cosmic messengers can enhance our comprehension of FIP phenomenology. These analyses set a new standard for the methodologies employed in this type of investigation and establish solid groundwork for future inquiries that aim to shed light on the characteristics of FIPs and a broad set of sub-GeV electron/positrons sources.

{\footnotesize \section*{Acknowledgments}
We thank Jordan Koechler, Marco Cirelli, Francesca Calore, Thomas Siegert, Leonardo Mastrototaro, Daniele Gaggero and Tim Linden for useful discussions. 
This article is based upon work from COST Action COSMIC WISPers CA21106, supported by COST (European Cooperation in Science and Technology).
The work of PC is supported by the European Research Council under Grant No.~742104 and by the Swedish Research Council (VR) under grants  2018-03641 and 2019-02337. 
PDL is currently supported by the Juan de la Cierva JDC2022-048916-I grant, funded by MCIU/AEI/10.13039/501100011033 European Union "NextGenerationEU"/PRTR. The work of PDL is also supported by the grants PID2021-125331NB-I00 and CEX2020-001007-S, both funded by MCIN/AEI/10.13039/501100011033 and by ``ERDF A way of making Europe''. 
SB is supported by the STFC under grant ST/X000753/1.
The work of PC is supported by the Swedish Research Council (VR) under grants  2018-03641 and 2019-02337.
PDL also acknowledges the MultiDark Network, ref. RED2022-134411-T.}
{\small
\section*{References}
\bibliography{biblio}}

\end{document}